\documentstyle[psfig, twoside]{article}
\def\bitem{\par\smallskip\noindent\hangindent12pt\small}

\def\phd0{\hbox{$\phantom{\hbox{.0}}$}}
\newcommand\apj{{ApJ}}%
\newcommand\aaps{{A\&AS}}%
\newcommand\pasp{{PASP}}%
\newcommand\ssr{{Space~Sci.~Rev.}}%
\newcommand{\Ang}{\AA }

\begin{document}
\markboth{R.A. Downes et al.}{A Catalog and Atlas of Cataclysmic Variables:
The Final Edition}

\title{\parbox{\textwidth}{ {\normalsize \sl 2005: The Journal of 
Astronomical Data 11, 2.   \hfill
\copyright{}~R.A. Downes et al.  }}\\ {\vspace{5mm} 
A Catalog and Atlas of Cataclysmic Variables:
The Final Edition}}

\author{Ronald A. Downes (1),
Ronald F. Webbink (2),
Michael M. Shara (3),\\
Hans Ritter (4),
Ulrich Kolb (5),~and
Hilmar W. Duerbeck (6)
\\
\small (1) Space Telescope Science Institute, 3700 San Martin Drive, 
Baltimore, MD 21218, USA \\
\small (2) Department of Astronomy, University of Illinois, 103 Astronomy
Building, \\
\small     1002 West Green Street, Urbana, IL 61801 \\
\small (3) American Museum of Natural History, Astrophysics Department, \\
\small     Central Park West and 79th Street, New York, NY 10024 \\
\small (4) Max-Planck-Institut f\"ur Astrophysik, \\
\small     Karl-Schwarzschild-Sterne 1, D-85741 Garching, Germany \\
\small (5) Department of Physics and Astronomy, The Open University, \\
\small     Walton Hall, Milton Keynes MK7 6AA, UK \\
\small (6) WE/OBSS, Free University (VUB), Pleinlaan 2, B-1050 Brussels, \\
\small     Belgium\\
}
\date{\small Received 5 December 2005; accepted 31 December 2005}

\maketitle
\vspace{-1cm}

\section*{Abstract}

The Catalog and Atlas of Cataclysmic Variables has been a staple
of the CV community for over 10 years.  The catalog has grown from
751 CVs in 1993 to 1600 CVs at present.  The catalog became
a ``living'' edition in 2001, and its contents have been continually
updated since that time.  Effective 27 January 2006, the catalog
will transition to an archival site, with no further updates to
its contents.  While it is antipicated that the site will remain
active, we present the complete contents of the site as a 
precaution against a loss of the on-line data.

\vspace{2mm}
\noindent {\bf Keywords:} catalogs:cataclysmic variables

\section{Introduction}

The Catalog and Atlas of Cataclysmic Variables (Edition 1: Downes and
Shara 1993; Edition 2: Downes, Webbink, and Shara 1997; Living
Edition: Downes et al.~2001) has been a prime source of information
for the cataclysmic variable (CV) community for over 10 years.  The
catalog has also grown substantially over time, from 751 CVs in the
first edition, to 865 CVs in the second edition, to 1600 CVs at
present.  Over time, it has evolved from a paper document to a
``living'' document, which allowed for rapid updates to its content.

Due to the retirement of the lead author, the on-line catalog will
transition from an active site to an archival site effective 27
January 2006, with no further updates to its content.  This paper
provides the final content of the catalog and atlas, although it is
planned for the site to remain accessible for the forseeable future.

Section 2 describes the catalog, while Section 3 describes the
Atlas. Section 4 summarizes the state of the catalog.

\section{The Catalog}

The catalog was frozen on 27 January 2006, and the final statistics
are given in Table 1.  Due to the size of the catalog, several files
are needed to supply all the information, and these files listed in
Table 2.

The fields in the catalog are:

\noindent
{\bf GCVS NAME} - the name of the object in the General Catalogue of
Variable Stars and subsequent Namelists. For those objects without
variable star designations, we list the constellation name only (which
were derived from Roman (1987)).  Since some constellations contain
more than one object without a GCVS designation, in previous versions
we included a number (a strictly provisional designation) after the
constellation name. In the on-line version of the catalog, such ad hoc
numbering was no longer required. However, for ease of comparison with
previous versions, we have retained this type of designation for all
objects that had it.

\noindent
{\bf COORDINATES} - whenever possible, the J2000 coordinates of the
objects as measured in the International Celestial Reference System
(ICRS), or taken from the literature (see Coordinate Reference). The
right ascension is given to the nearest 0.01s,
while the declination is given to the nearest 0.1s; for objects that
are very faint or not visible, the coordinates are given to a lesser
accuracy.  For faint novae, the coordinates are taken from Duerbeck
(1987) or from the literature (for recent novae), and are precessed to
the J2000 equinox. For those objects without available/usable finding
charts, coordinates have been obtained from the literature, and are
generally given to a lesser accuracy than those objects measured in
this work.

\noindent
{\bf PROPER MOTION} - the proper motion for the object in
arcseconds/yr (RA and Dec), along with the associate errors. The epoch
for the coordinates is also given.  Users should consult the Proper
Motion Reference for the definition of the corresponding reference
frame.

\noindent
{\bf GALACTIC COORDINATES} - the galactic longitude (l) and latitude (b).

\noindent
{\bf TYPE} - the type of variability of the object. Table~3 lists
the various types used in this work, which is based on the
classification scheme used in the GCVS. Those types in upper-case
letters are taken directly from the GCVS, while those in lower-case
letters have been obtained from the literature; these are generally in
agreement with the GCVS and are presumably more reliable.  Whenever
possible, a type from the literature was used. 
There are many objects in the catalog designated NON-CV, which are
stars that have been previously cataloged as CVs, and are included for
completeness; the references for these stars are those papers which
refute the CV nature of the objects. The revised classification is
given in the notes for each object.

\noindent
{\bf PERIOD} - the orbital period (in days) for the object.

\noindent
{\bf YEAR OF OUTBURST} - the year of outburst (for novae).

\noindent
{\bf CLUSTER SOURCE} - a flag indicating if the object is in a globular (G)
or open (O) cluster.

\noindent
{\bf MAGNITUDE RANGE} - the MAXimum and MINimum magnitudes for the
objects; the magnitude systems are listed in the Table~4. For novae,
the primary sources are Duerbeck (1987) and Duerbeck (2001,
unpublished), while for the non-novae, the catalog of Ritter and Kolb
(1998 and later updates) is the prime source.  When no other
references to brightness were available, the GCVS values are used.

\noindent
{\bf COORDINATE REFERENCE} - a code for a reference to the
coordinates. Entries listed as ICRS (International Celestial Reference
System) have been measured by the authors in that reference frame;
other codes refer to references from the literature (see the
References.html file for the code definitions).  An asterisk (*)
following the reference means that there is a comment regarding the
coordinate measurement (see the ASCII Notes Report for the object).

\noindent
{\bf PROPER MOTION REFERENCE} - a code for a reference to the proper motion
(see the References.html file for the code definitions).  

\noindent
{\bf TYPE REFERENCE} - a code for a reference to the CV classification (see
the References.html file for the code definitions).  An asterisk (*)
following the reference means that there is a comment regarding the
classification (see the ASCII Notes Report for the object).

\noindent
{\bf CHART REFERENCE} - a code for the original reference from which our
chart is based (see the References.html file for the code
definitions).  Note that the identifications of the CVs are based on
the published charts (or in some cases coordinates only), and have not
been independently verified by the authors. An asterisk (*) following
the references means that there is a comment regarding the
identification (see the ASCII Notes Report for the object).
    
\noindent
{\bf SPECTRUM REFERENCE} - a code for a reference to a published
spectrum (see the References.html file for the code definitions).  A
suffix S indicates a spectrum in quiescence, while a suffix of X
indicates a spectrum in outburst. Whenever available, the quiescent
spectrum was chosen for the catalog over an outburst spectrum. A colon
(:) following the reference indicates that the spectrum is only
described, the spectrum is a glass plate tracing, or the reference is
unconfirmed (for only a few novae in Duerbeck's atlas).  An asterisk
(*) following the references means that there is a comment regarding
the spectrum (see the ASCII Notes Report for the object).

\noindent
{\bf PERIOD REFERENCE} - a code for a reference to the period (see the
References.html file for the code definitions).  Note that all periods
come from the catalog of Ritter and Kolb (1998 and later updates), or
from Ritter (private communication).

\noindent
{\bf OTHER NAME} - discovery or common alternative (non-GCVS) designation
for the object.

\noindent
{\bf SPACE-BASED OBSERVATIONS} - a "Y" in the field means that data from
that satellite exists; for HST data, the notes for the object indicate
if the data are imaging, spectroscopy, photometry, and/or astrometric.

\section{The Atlas}

Finding charts for all objects with chart references (including
objects which we have identified based on positional coincidence) are
included (see the Finding Charts directory).  Most charts are based on
the Digitized Sky Survey, and since those plates vary in both color
and limiting magnitude, we note (see the Finding Charts details file)
the emulsion and exposure time for each chart. The field-of-view for
the charts is also indicated (mostly 5' x 5').  First Generation DSS
emulsion/filter combinations are defined in the Table 5, while the
Second Generation DSS emulsion/filter combinations are defined in
Table 6.

Some charts are ground-based CCD images, while others (mostly globular
cluster CVs) are HST images; the filters and exposure times for these
are given in the Finding Charts details file.  North is up and East to
the left for all charts unless explicitly noted (for some HST images).
Tick marks are used to identify the object, while circles are used for
those fields where this is not a definitive identification.  Note that
for some extremely crowded fields, a circle has been used to mark the
object.

\section{Summary}

The Catalog and Atlas of Cataclysmic Variables ``living'' web
site will transition to an archival site on 27 January 2006.
Although the site should remain accessible after that date, we
provide the complete contents of the catalog and atlas as a 
precaution against a loss of the archival data.

\vspace{-.2cm}
\section*{Acknowledgments}

We thank all the astronomers, both professional and amateur, who
provided information to support this catalog.  We thank Sarah
Stevens-Rayburn and Brenda Corbin for their excellent assistance in
obtaining some of the more obscure references needed to generate the
finding charts, and Mike Potter, Debra Wallace, and Matt McMaster for
assistance in measuring the coordinates and generating the finding
charts.  We also thank Anne Gonnella, Steve Hulbert, Calvin Tullos,
and Mike Wiggs for the work in creating the web site.  R.F.W.
acknowledges the support from NSF grants AST 9618462 and AST 0406726.

\newpage
\section*{References}

\bitem Downes, R.A. and Shara, M.M. 1993, \pasp, 105, 127 
\bitem Downes, R.A., Webbink, R.F., and Shara, M.M. 1997, \pasp, 109, 345 
\bitem Downes, R.A., Webbink, R.F., Shara, M.M., Ritter, H., Kolb, U. and
                     Duerbeck, H.W. 2001, \pasp, 113, 764
\bitem Duerbeck, H.W. 1987, \ssr, 45, 1
\bitem Ritter, H. and Kolb, U. 1998, \aaps, 129, 83
\bitem Roman, N.G. 1987, \pasp, 99, 695
\bitem Steiner, J.E. and Diaz, M.P. 1998, \pasp, 110, 276
\bitem Webbink, R.F., Livio, M., Truran, J.W., and Orio, M. 1987, \apj, 314, 653

\scriptsize{

\begin{table}

Table 1. CV Catalog Statistics
\vspace*{1mm}

\begin{tabular}{lr}\hline
Object Type & Number of Objects \\
\hline
Number of Objects in the Catalog & 1829 \\
Number of CV's in the Catalog & 1600 \\
Number of non-CVs in the catalog & 229 \\
Number of Globular Cluster CVs in the catalog & 158 \\
Number of Open Cluster CVs in the catalog & 6 \\
\\
Number of UG in the catalog & 636 ( 40\%) \\
Number of novae in the catalog & 337 ( 21\%) \\
Number of NL in the catalog & 263 ( 16\%) \\
Number of cvs in the catalog & 342 ( 21\%) \\
Number of ibwd in the catalog & 17 ( 1\%) \\
Number of cbss in the catalog & 3 ( 0\%) \\
Number of SNe in the catalog & 1 ( 0\%) \\
Number of cvs with no GCVS name & 716 ( 45\%) \\
\\
Number of UGs with quiescent spectra & 300 ( 47\%) \\
Number of UGs with outburst spectra & 37 ( 6\%) \\
Number of UGs with no spectra & 299 ( 47\%) \\
\\
Number of novae with quiescent spectra & 77 ( 23\%) \\
Number of novae with outburst spectra & 148 ( 44\%) \\
Number of novae with no spectra & 112 ( 33\%) \\
\\
Number of novalikes with quiescent spectra & 223 ( 85\%) \\
Number of novalikes with no spectra & 40 ( 15\%) \\
\\
Number of cvs with quiescent spectra & 158 ( 46\%) \\
Number of cvs with outburst spectra & 5 ( 1\%) \\
Number of cvs with no spectra & 179 ( 52\%) \\
\\
Number of IBWDs with quiescent spectra & 16 ( 94\%) \\
Number of IBWDs with outburst spectra & 1 ( 6\%) \\
\\
Number of CBSSs with quiescent spectra & 2 ( 67\%) \\
Number of CBSSs with no spectra & 1 ( 33\%) \\
\\
Number of objects with quiescent spectra & 776 ( 49\%) \\
Number of objects with outburst spectra & 191 ( 12\%) \\
Number of objects with no spectra & 632 ( 40\%) \\
\\
Number of objects with a known period & 477 ( 30\%) \\
Number of objects with a known period $<1$ hour & 14 ( 3\%) \\
Number of objects with a known period between 1-2 hours & 159 ( 33\%) \\
Number of objects with a known period between 2-3 hours & 52 ( 11\%) \\
Number of objects with a known period between 3-4 hours & 103 ( 22\%) \\
Number of objects with a known period between 4-5 hours & 50 ( 10\%) \\
Number of objects with a known period between 5-6 hours & 25 ( 5\%) \\
Number of objects with a known period between 6-7 hours & 20 ( 4\%) \\
Number of objects with a known period between 7-8 hours & 10 ( 2\%) \\
Number of objects with a known period between 8-9 hours & 9 ( 2\%) \\
Number of objects with a known period $>9$ hours & 34 ( 7\%) \\
 \\
Number of CVs with significant ($>3\sigma$) proper motions & 433 \\
\\
Number of objects with IUE data & 192 \\
Number of objects with HST data & 289 \\
Number of objects with Ariel 5 data & 8 \\
Number of objects with HEAO-1 data & 42 \\
Number of objects with HEAO-2 data & 67 \\
Number of objects with EXOSAT data & 83 \\
Number of objects with ROSAT data & 297 \\
Number of objects with Ginga data & 20 \\
Number of objects with ASCA data & 61 \\
Number of objects with EUVE data & 35 \\

\end{tabular}
\end{table}

\clearpage

\begin{table}

Table 2. Files included in this publication
\vspace*{1mm}

\begin{tabular}{ll}\hline
Name & Contents \\
\hline
References & a html file which defines all the reference codes \\
ASCII Report & a html file listing all the object data, sorted by
right ascension \\
ASCII Report, comma separated & a text file (csv format) listing all the object data,
sorted by right ascension \\
ASCII Notes Report & a html file list all the object notes, sorted by
right ascension \\
Finding Chart details & a text file listing the information (image source,
exposure time, and field size) for each chart \\
Finding Charts & a directory containing jpg images of the finding charts \\

\end{tabular}
\end{table}

\begin{table}

Table 3. CV Classifications
\vspace*{1mm}

\begin{tabular}{ll}\hline
Type & Definition\\
\hline

UG     & U Gem variable (dwarf nova)   \\
UGZ    & U Gem variable (Z Cam subtype)   \\    
UGSS   & U Gem variable (SS Cyg subtype)   \\
UGSU   & U Gem variable (SU UMa subtype)      \\
UGWZ   & U Gem variable (WZ Sge subtype)      \\
  \\
N      & nova   \\
NA     & fast nova \\  
NB     & slow nova   \\
NC     & very slow nova \\  
NR     & recurrent nova   \\
NRA    & recurrent nova - giant donor (Webbink et al. 1987)   \\
NRB    & recurrent nova - non-giant donor (Webbink et al. 1987) \\
SN     & possible supernova with no galaxy visible   \\
  \\
NL     & novalike variable   \\
NLV    & novalike variable (V Sge subtype; Steiner and Diaz 1998) \\
UX     & novalike variable (UX UMa subtype)   \\
VY     & novalike variable (VY Scl subtype - systems which undergo low states)   \\
AM     & AM Herculis variable (synchronous rotators)   \\
DQ     & DQ Herculis variable (non-synchronous rotators)  \\ 
CV     & cataclysmic variable (no type classification)   \\
CBSS   & close binary supersoft x-ray source \\
     \\ 
IBWD   & interacting binary white dwarf   \\
  \\
M      & Mira variable   \\
I      & Irregular variable \\  
UV     & UV Ceti-type star   \\
Z AND  & symbiotic variable (Z And subtype) \\  
NON-CV & not a cataclysmic variable (although once classified as such)   \\
NON-EX & non-existent object   \\
  \\
PEC    & peculiar   \\
:      & uncertain   \\
::     & very uncertain \\  

\end{tabular}
\end{table}

\begin{table}

Table 4. Magnitude Systems
\vspace*{1mm}

\begin{tabular}{cl}\hline
Magnitude Code & Definition \\
\hline

   U     & Johnson U \\
   B     & Johnson B \\
   V     & Johnson V \\
   R     & Johnson R \\
   I     & Johnson I \\
   c     & unfiltered CCD \\
   f     & m(2200\Ang)\\
   g     & Gunn g \\
   h     & HIPPARCOS magnitude system \\
   i     & Gunn i \\
   j     & SRC j (unfiltered IIIa-J) \\
   p     & photographic \\
   r     & red \\
   s     & Sloan g*\\
   u     & m(1400\Ang)\\
   v     & visual \\
   w     & m(3000\Ang)\\
   x     & m(F336W)\\

\end{tabular}
\end{table}

\begin{table}

Table 5. First Generation DSS Emulsion Codes 
\vspace*{1mm}

\begin{tabular}{cl}\hline
Emulsion Code & Definition \\
\hline

POSSI-E  & POSS-E RED PLATE \\
XV       & SERC-V Equatorial extension \\
S        & SERC-J Survey \\

\end{tabular}
\end{table}

\begin{table}

Table 6. Second Generation DSS Emulsion Codes 
\vspace*{1mm}

\begin{tabular}{cl}\hline
Emulsion Code & Definition \\
\hline

UK-F     & 'Galactic Red' survey (UK Schmidt) IIIaF + RG610 SHORT exposure in galactic plane \\
POSSII-F & POSS-II Red IIIaF + RG610 \\
POSSII-J & POSS-II Blue IIIaJ + GG385 \\
ER       & 'Equatorial Red' survey (UK Schmidt) IIIaF + RG610 \\
XS       & 'Second Epoch Southern' survey (UK Schmidt) IIIaF + RG610 \\

\end{tabular}
\end{table}

\end{document}